\documentclass[superscriptaddress, twocolumn, prd, showpacs, nofootinbib,
a4]{revtex4-1}
\usepackage{graphicx}
\usepackage{enumerate}
\usepackage{tikz}
\usepackage{amsmath}
\usepackage{amssymb}
\usetikzlibrary{arrows,positioning}

\begin{document}

\title{New limits on dark matter annihilation from AMS cosmic ray positron data}

\author{Lars Bergstr\"om}
\email{lbe@fysik.su.se}
\affiliation{The Oskar Klein Centre for Cosmoparticle Physics, Department of
Physics, Stockholm University, AlbaNova, SE-106 91 Stockholm, Sweden}

\author{Torsten~Bringmann}
\email{torsten.bringmann@desy.de}
\affiliation{{II.} Institute for Theoretical Physics, University of Hamburg,
Luruper Chausse 149, DE-22761 Hamburg, Germany}

\author{Ilias Cholis}
\email{cholis@fnal.gov}
\affiliation{Center for Particle Astrophysics, Fermi National Accelerator
Laboratory, Batavia, IL 60510,USA}

\author{Dan Hooper}
\email{dhooper@fnal.gov}
\affiliation{Center for Particle Astrophysics, Fermi National Accelerator
Laboratory, Batavia, IL 60510,USA}
\affiliation{University of Chicago, Department of Astronomy and Astrophysics,
Chicago, Illinois, 60637, USA}

\author{Christoph Weniger}
\email{c.weniger@uva.nl}
\affiliation{GRAPPA Institute, University of Amsterdam, Science Park 904, 1090
GL Amsterdam, Netherlands}

\date{October 25, 2013}

\begin{abstract}
  The AMS experiment onboard the International Space Station has recently
  provided cosmic ray electron and positron data with unprecedented precision
  in the range from 0.5 to 350\,GeV. The observed rise in the positron fraction
  at energies above 10\,GeV remains unexplained, with proposed  solutions
  ranging from local pulsars to TeV-scale dark matter. Here, we make use of this 
  high quality data to place stringent limits on dark matter with masses below 
  $\sim$300\,GeV, annihilating or decaying to leptonic final states, essentially
  independent of the origin of this rise. We significantly improve on 
  existing constraints, in some cases by up to two orders of magnitude.
\end{abstract}

\pacs{95.85.Ry, 95.35.+d, 95.30.Cq; FERMILAB-PUB-13-202-A}

\maketitle

\paragraph*{Introduction.}
The AMS (Alpha Magnetic Spectrometer) collaboration has very recently 
announced the results of its first data collected from the International Space 
Station \cite{ams}, consisting of a high precision measurement of the cosmic
ray (CR) positron fraction \cite{Aguilar:2013qda}. This new data provides a 
confirmation of the rise of this quantity above 10\,GeV, as previously observed by 
PAMELA \cite{Adriani:2008zr} and Fermi \cite{FermiLAT:2011ab} (and with earlier 
hints provided by HEAT \cite{Barwick:1997ig} and AMS-01 \cite{Aguilar:2007yf}).  
Such a rise is not predicted in the standard scenario, in which CR positrons are 
mostly produced as {\it secondary} particles, as a result of collisions of CR protons 
with the interstellar medium (ISM). Instead, the large positron fraction seems to 
require the existence of at least one additional nearby {\it primary} source of high 
energy positrons. Local pulsars have emerged as the leading
astrophysical candidates \cite{Hooper:2008kg,Profumo:2008ms}, although it has
also been argued that strong local sources might not actually be needed when
taking into account the spiral structure of the Milky Way in full 3-D
propagation models \cite{Gaggero:2013rya} and that even a secondary production
mechanism in the shock waves of supernovae remnants 
\cite{Blasi:2009hv,Mertsch:2009ph} could provide a viable mechanism to explain 
the data  \cite{Serpico:2011wg}.

A more exotic possibility is that the observed positrons may be produced in the 
annihilations or decays of TeV-scale dark matter (DM) particles. Such scenarios, 
however, require unexpectedly large annihilation rates into predominantly leptonic 
final states \cite{Bergstrom:2009fa, Finkbeiner:2010sm, Yuan:2013eja,
Cholis:2013psa, Jin:2013nta} and are subject to significant constraints from CR
antiproton, gamma-ray and synchrotron data \cite{Donato:2008jk,
Bertone:2008xr, Bergstrom:2008ag, Cirelli:2009dv, Evoli:2011id,Kappl:2011jw,
Wechakama:2012yb, DeSimone:2013fia}.  Upcoming
AMS data may help to settle this open issue not only by increasing statistics and 
extending their study to higher energies, but also by providing high precision 
measurements of other CR particle spectra (likely breaking degeneracies in the 
propagation parameters \cite{breakdegen}). Fermi and AMS will also further 
constrain any anisotropy in the positron/electron flux (where current limits are already
close to discriminating between some of the scenarios described above
\cite{Ackermann:2010ip,DiBernardo:2010is}).

In this Letter, we do not make any attempt to explain the origin of the rise in the
positron fraction. Instead, we focus on using the AMS data
to derive limits on subdominant exotic contributions to the observed 
CR positron spectrum, in particular from DM with masses below $\sim$300\,GeV. 
While positrons have been used in the past to probe DM annihilation or 
decay~\cite{Ellis:1988qp,Rudaz:1987ry,Kamionkowski:1990ty,Baltz:1998xv,
Baltz:2001ir,Kopp:2013eka}, we exploit here for the first time the extremely 
high quality of the AMS data to search for pronounced {\it spectral features} in the 
positron flux predicted in some DM models 
\cite{Tylka:1989xj, Turner:1989kg,Baltz:2002we,Hooper:2004xn,Baltz:2004ie,
Bergstrom:2008gr,Hooper:2012gq}. Much as exploiting spectral features can 
significantly improve the sensitivity of indirect DM searches using gamma 
rays~\cite{Bringmann:2011ye}, we demonstrate that the same is true 
for positrons, despite energy losses and other complicating factors.  We derive 
limits that exceed the currently most stringent results on DM annihilation into 
leptons \cite{Ackermann:2011wa, cmb} by up to two orders of magnitude.

This Letter is organized as follows. We first briefly review various
astrophysical sources of  leptons and how they manifest themselves in the
observed CR  flux, and then discuss possible contributions from DM. We continue 
with a description of the statistical treatment implemented here, before moving on
to present our main results and conclusions. In an Appendix \cite{supp}, we collect
further technical details of our procedure for deriving limits on a possible
DM signal, discussing in particular the impact of systematic uncertainties in
the background modeling.

\paragraph*{Astrophysical origins of cosmic ray leptons.}

The origin of high energy electrons can be traced back to i) supernova explosions
that accelerate the ISM to produce what are typically referred to as primary CRs, ii)
inelastic collisions of primary CR protons and nuclei with the ISM
(resulting in charged mesons, which decay, producing secondary electrons
{\it and} positrons), and  iii) individual sources such as  pulsars that
produce $e^\pm$ pairs. The averaged spectrum of propagated primary CR 
electrons (originating from many supernovae) is expected to be harder than that 
of the secondary $e^{\pm}$ component because the primary CR progenitors of 
the secondaries have also experienced propagation effects;
both spectra are well described by power-laws, with spectral indices of about
$3.3$ to $3.5$ ($3.7$) for  primary electrons (secondary $e^{\pm}$) at energies 
above $\sim$10\,GeV \cite{secondaries}.  The contribution from all galactic pulsars can be 
approximated by a  power-law with an exponential cut-off at high energies, with a 
propagated spectral index of 2.0 $\pm$ 0.5 \cite{Hooper:2008kg, Profumo:2008ms}.

The Galactic Magnetic Field at  scales $\gtrsim100$\,pc has a random and a regular
component \cite{Jansson:2009ip}. As CR leptons propagate away from
their sources, they follow the field lines and scatter off $B$-field
irregularities.  The net effect can be approximated as a random walk diffusion
within a zone surrounding the Galactic Disk \cite{LongairDiff, Strong:2007nh}.
Further away from the disk the magnetic fields become weak,
essentially leading to freely propagating CRs. During their propagation throughout 
the Galaxy, electrons and positrons also experience significant energy losses due to 
synchrotron and inverse Compton scattering on the galactic radiation field and the 
cosmic microwave background. The impact of other
effects such as convective winds, ionization losses, or positron annihilation in
collisions with matter are not significant for leptons in the energy range considered 
here \cite{Strong:2007nh, LongairEloss} and are therefore ignored. We do, however, 
include bremsstrahlung emission, diffusive re-acceleration, and solar
modulation  inside the heliosphere (using the force-field  approximation
\cite{Gleeson1968}), which have an impact on CR $e^{\pm}$ spectra below
5--10\,GeV \cite{Strauss2012, Maccione:2012cu}.

For the propagation of CR leptons, we use the standard numerical tool 
GALPROP v54 \cite{galprop}, which includes up-to-date  implementations of  
 the local interstellar radiation field and galactic gas distribution. These are 
relevant for both the production of secondary leptons and energy losses.  GALPROP 
assumes a diffusion zone with cylindrical 
symmetry within which CRs diffuse and beyond which they escape. Its scale height,
$L$, and other diffusion parameters, notably the diffusion time-scale and local diffusion
properties, are constrained by observed CR ratios, including $\bar{p}/p$, B/C and
$^{10}$Be/$^9$Be.
 As reference values we assume $L=4$ kpc, corresponding to
the value best-fit by CR data \cite{Trotta:2010mx}  and favored by radio observations 
\cite{Bringmann:2011py}, and the standard default GALPROP assumptions for the 
local radiation and magnetic field energy densities, corresponding to 
$U_{\rm rad} + U_{B} = 1.7$\,eV\,cm$^{-3}$ \cite{galprop}.  For the diffusion zone 
scale height, values of $L<2$\,kpc are in tension with a combined analysis of CR 
and  gamma-ray data \cite{Cholis:2011un}, while increasing $L$ beyond $8$\,kpc 
does not significantly alter our results.\footnote{For $L=$ 2,
4, 8 kpc and rigidity $R$, we adopt a diffusion coefficient $D( R)= D_{0}
(\frac{R}{1 GV})^{0.5}$, with $D_{0} =$ 0.81, 1.90, 2.65 
($\times 10^{28}$ cm$^{2}$s$^{-1}$), and  Alfv\'en velocities  9, 10, 10 km s$^{-1}$.}
The propagation of high-energy leptons is actually dominated by energy losses 
rather than diffusion, implying that more conservative limits would arise for larger 
values of the local radiation and
magnetic field energy densities. In our subsequent discussion, we will allow
for an increase of $U_{\rm rad} + U_{B}$ by up to $50\%$ with respect to the
reference value, which is still compatible with gamma-ray and synchrotron data
\cite{Jaffe:2011qw, Bringmann:2011py}.

\paragraph*{Positrons from dark matter.}

DM particles annihilating or decaying in the Galactic Halo may also contribute to 
the CR lepton spectrum, producing equal numbers of positrons and 
electrons. For annihilating DM, the injected spectrum of CR leptons per volume and 
time  is given by $Q=\frac12\langle\sigma v\rangle\left(\rho_\chi/m_\chi\right)^2 dN/dE$ 
(divided by 2 if the DM particle is not self-conjugate), while for decaying DM, 
this is instead $Q=\Gamma\rho_\chi/m_\chi dN/dE$, where $\Gamma$ is the decay 
rate. Here, $\langle\sigma v\rangle$ is the velocity-averaged annihilation rate, 
$\rho_{\chi}$ is the 
DM density, $m_{\chi}$ is the DM mass, and $dN/dE$ is 
the spectrum of leptons produced per annihilation or decay. As our default choice, 
we adopt a DM distribution which follows an Einasto profile
\cite{Merritt:2005xc},
normalized to a local density of $\rho^\odot_\chi=0.4$\,GeV 
\cite{Catena:2009mf,Salucci:2010qr}.

Positrons from DM annihilation or decay typically result from the decay of $\pi^+$ 
(for hadronic final states), or the leptonic decay of $\tau^+$ or $\mu^+$. 
Owing to the high multiplicity of such processes, the resulting $e^+$ energy
distribution at injection (which we take from Ref.~\cite{Cirelli:2010xx}) is typically very soft. If  DM annihilates directly into 
$e^\pm$, however, these are produced nearly monochromatically.
Even  after accounting 
for energy losses from propagation, a very characteristic spectrum 
arises in this case, with a sharp edge-like feature at $E\!=\!m_\chi$ 
(or at $E\!=\!m_\chi/2$ for decaying
DM). A comparably distinct spectral feature arises from the annihilation of
Majorana DM into $e^+e^-\gamma$ final states.  Popular examples for DM 
models with large annihilation rates into $e^\pm$ final states include Kaluza-Klein
DM \cite{Hooper:2004xn}, while the supersymmetric neutralino is a possible
candidate for producing a spectrum dominated by $e^+e^-\gamma$ final states
\cite{Bergstrom:2008gr}.\footnote{ 
By $e^+e^-\gamma$  we will always refer to
that specific situation, dominated by photon emission off virtual
selectrons $\tilde e$.  
We assume at least one of the $\tilde e$ to be degenerate in mass with
the neutralino.}

\begin{figure}[t!]
\includegraphics[width=\linewidth]{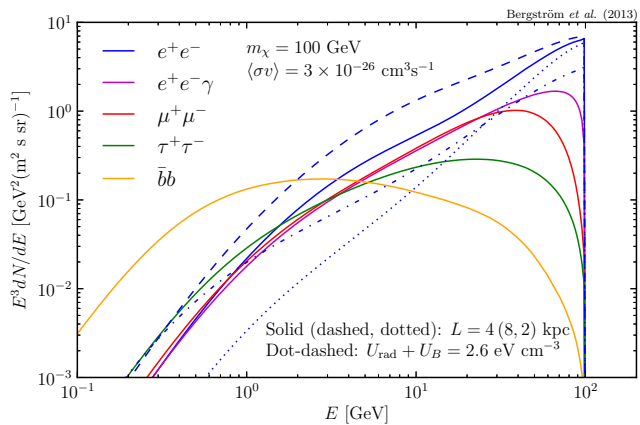}
  \caption{The $e^{\pm}$ spectrum from annihilating DM, after propagation, for 
    different annihilation
    final states, assuming  $\langle \sigma v\rangle$$=3\times 10^{-26}$
    cm$^{3}$s$^{-1}$.  Solid lines refer to
    reference diffusion zone ($L$=4\,kpc) and energy loss assumptions ($U_{\rm
    rad}+U_{B} = 1.7$\,eV\,cm$^{-3}$). Dashed (dotted) lines show the effect
    of a different  scale height  $L$=8\,(2) kpc. The dash-dotted line shows the 
    impact of increasing the local radiation plus magnetic field density to 
    $U_{\rm rad}+U_{B} = 2.6$\,eV\,cm$^{-3}$.}
  \label{fig:spectra}
\end{figure}

\begin{figure}[t!]
  \includegraphics[width=\linewidth]{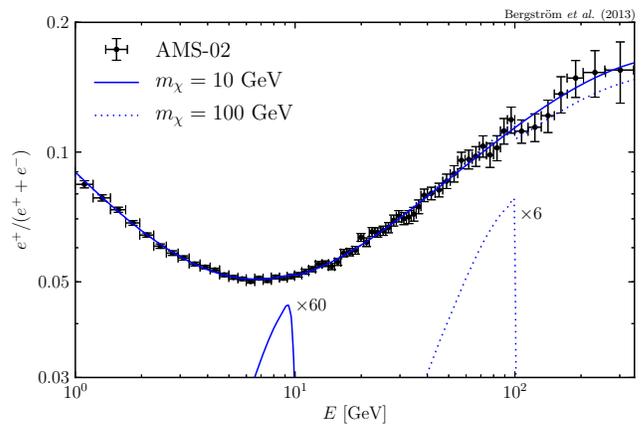}
  \caption{The AMS positron fraction measurement \cite{Aguilar:2013qda} and 
  background+signal fit for DM annihilating directly to $e^+ e^-$, for 
  $m_\chi=10$\,GeV and 100\,GeV. The normalization of the DM signal in 
  each case was chosen such that it is barely excluded
    at the $95\%$ CL.   For better visibility, the contribution from DM
    (lower lines) has been rescaled as indicated.}
  \label{fig:fraction}
\end{figure}

We illustrate this in Fig.~\ref{fig:spectra} by showing the propagated
$e^\pm$ spectra for various  final states and an annihilation rate that
corresponds to the ``thermal" cross section of $\langle \sigma v \rangle_{\rm
therm} \equiv 3\times 10^{-26}\rm \ cm^3 s^{-1}$ (which leads to the correct
relic density in the simplest models of thermally produced DM). As
anticipated, the $e^+e^-$ and $e^+e^-\gamma$ final states result in the most 
pronounced spectral features
-- a fact which helps considerably, as we will see, to distinguish them from
astrophysical backgrounds. For the case of $e^+e^-$ final states, we also show
how the spectrum depends on our local diffusion and energy loss assumptions
within the range discussed above. Increasing $L$ enables CR leptons to reach 
us from greater distances due to the larger diffusion volume and therefore
results in softer propagated spectra. While the peak normalization of the
spectrum depends only marginally on $L$, it may be reduced by up to a factor
of $\sim$2 when increasing the assumed  local energy losses via synchrotron
radiation and inverse Compton scattering by 50\%.
In Fig.~\ref{fig:fraction}, we show a direct
comparison of the DM signal with the AMS data, for the case of $e^+e^-$ final
states contributing at the maximum level allowed by our constraints (see
below) for two fiducial values of $m_\chi$.  Again, it should be obvious that
the shape of the DM contribution differs at all energies significantly from
that of the background.

\paragraph*{Statistical treatment.}
We use the likelihood ratio test~\cite{Rolke:2004mj} to determine the
significance of, and limits on, a possible DM contribution to the positron
fraction measured by AMS.  As likelihood function, we adopt a product of
normal distributions $\mathcal{L}=\prod_i N(f_i|\mu_i, \sigma_i)$; $f_i$ is
the measured value, $\mu_i$ the positron fraction predicted by the model, and
$\sigma_i$ its variance. The DM contribution enters with a single degree of
freedom, given by the non-negative signal normalization. Upper limits at the
$95\%$CL on the DM annihilation or decay rate are therefore derived by
increasing the signal normalization from its best-fit value until
$-2\ln\mathcal{L}$ is changed by 2.71, while profiling over the parameters of 
the background model.

We use data in the energy range 1--350\,GeV; the variance $\sigma_i$ is
approximated by adding the statistical and systematic errors of the
measurement in quadrature, $\sigma_i=(\sigma_{i,\rm stat}^2 + \sigma_{i,\rm
sys}^2)^{1/2}$.  Since the total relative error is always small (below 17\%),
and at energies above 4\,GeV dominated by  statistics, we expect this
approximation to be very reliable.  The binning of the published positron
fraction follows the AMS energy resolution, which varies between 10.4\% at
1\,GeV and 1.5\% at 350\,GeV. Although we do not account for the finite energy 
resolution of AMS in our analysis, we have explicitly checked that this impacts 
our results by no more than 10\%. 

As our nominal model for the part of the $e^\pm$ spectrum that does not originate 
from DM, henceforth simply referred to as the astrophysical background, we  use
the same phenomenological parameterization as the AMS collaboration in  their
analysis \cite{Aguilar:2013qda}. This parameterization describes each of the 
$e^\pm$ fluxes as the sum of a common source spectrum -- modeled as a 
power-law with exponential cutoff -- and an individual power-law contribution 
(only the latter being
different for the $e^+$ and $e^-$ fluxes). After adjusting normalization and
slope of the secondary positrons such that the overall flux reproduces the
Fermi $e^+\!+\!e^-$ measurements  \cite{Ackermann:2010ij}, 
the five remaining
model parameters are left unconstrained. This phenomenological
parameterization provides an extremely good fit (with a $\chi^2/{\rm d.o.f.} =
28.5/57$), indicating that no fine structures are observed in the AMS data. For the 
best-fit spectral slopes of the individual power-laws we find 
$\gamma_{e^-} \simeq 3.1$ and $\gamma_{e^+} \simeq 3.8$, respectively, and 
for the common source $\gamma_{e^\pm} \simeq 2.5$ with a cutoff at 
$E_c\simeq$800\,GeV, consistent with Ref.~\cite{Aguilar:2013qda}. 
Subsequently, we will keep $E_c$ fixed to its best-fit value.

\paragraph*{Results and Discussion.}

\begin{figure}[t!]
  \begin{center}
    \includegraphics[width=\linewidth]{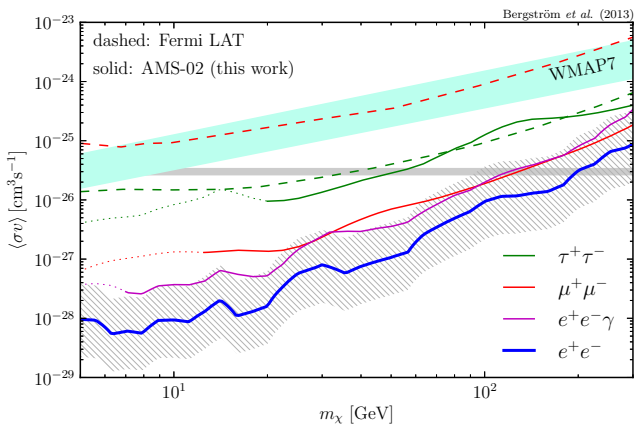}
  \end{center}
  \caption{Upper limits ($95\%$ CL) on the 
  DM annihilation cross section, as derived from the
    AMS positron fraction, for various final states (this work), WMAP7 
    (for $\ell^+\ell^-$) \cite{cmb} and Fermi LAT dwarf spheroidals (for $\mu^+ \mu^-$ 
    and $\tau^+ \tau^-$) \cite{Ackermann:2011wa}.  The dotted portions of the curves 
    are potentially affected by solar modulation. We also indicate $\langle \sigma v
    \rangle_{\rm therm} \equiv 3\times 10^{-26}\rm \ cm^3 s^{-1}$. The AMS
    limits are shown for reasonable reference values of the local DM density
    and energy loss rate (see text), and can vary by a factor of a few, as indicated by the
    hatched band (for clarity, this band is only shown around the $e^+ e^-$ constraint).}
    \label{fig:results}
\end{figure}

Our main results are the bounds on the DM annihilation cross section, as shown 
in Fig.~\ref{fig:results}.  No significant excess above background was observed.  For 
annihilations proceeding entirely to $e^+e^-$ final states, we find that the ``thermal'' 
cross section is firmly excluded for $m_\chi\lesssim 90$\,GeV. For $m_\chi\sim10$\,GeV, 
which is an interesting range in light of recent results from direct \cite{Bernabei:2010mq,
Aalseth:2011wp, Angloher:2011uu, Agnese:2013rvf, Frandsen:2013cna} and indirect 
\cite{Hooper:2011ti,Hooper:2013rwa,Abazajian:2012pn} DM searches, our upper 
bound on the annihilation cross section to $e^+ e^-$ is approximately two orders of 
magnitude below $\langle \sigma v\rangle_{\rm therm}$. If only a
fraction $f$ of DM annihilates like assumed, limits would scale like $f^{-2}$ (and, very roughly,
$\langle\sigma v\rangle_{\rm therm}\propto f^{-1}$).
We also show in Fig.~\ref{fig:results} the upper bounds obtained for  
other leptonic final states. As expected, these limits are weaker than those found in 
the case of direct annihilation to electrons -- both because part of the
energy is taken away by other particles (neutrinos, in particular) and because
they feature broader and less distinctive spectral shapes. These new limits on DM 
annihilating to $\mu^+ \mu^-$ and $\tau^+ \tau^-$ final states are still, however, 
highly competitive with or much stronger than those derived from other observations, 
such as from the cosmic microwave background \cite{cmb} and from gamma-ray 
observations of dwarf galaxies \cite{Ackermann:2011wa}. Note that for the case of
$e^+e^-\gamma$ final states even stronger limits can be derived for 
$m_\chi\gtrsim50$\,GeV by a spectral analysis of gamma
rays~\cite{Bringmann:2012vr}.
We do not show results for the $\bar b b$ channel, for which we nominally find even 
weaker limits due to the broader spectrum (for $m_\chi\simeq100$\,GeV, about
$\langle\sigma v\rangle\lesssim 1.1\cdot 10^{-24}\,\rm cm^3 s^{-1}$). In fact, due to degeneracies with the 
background modeling, limits for annihilation channels which produce such a broad 
spectrum of positrons can suffer from significant systematic uncertainties.  For this 
reason, we consider our limits on the $e^+e^-$ channel to be the most robust.

Uncertainties in the $e^{\pm}$ energy loss rate and local DM density weaken, to 
some extent, our ability to robustly constrain the annihilation cross sections under 
consideration in Fig.~\ref{fig:results}. We reflect this uncertainty by showing a band 
around the $e^+e^-$ constraint, corresponding to the range 
$U_{\rm rad} + U_{B} = (1.2-2.6)$\,eV\,cm$^{-3}$, and 
$\rho_\chi^\odot=(0.25-0.7)$\,GeV\,cm$^{-3}$~\cite{Salucci:2010qr, Iocco:2011jz}
(note that the \emph{form} of the DM profile has a much smaller impact).  
Uncertainty bands of the same width apply to each of the other final states shown 
in the figure, but are not explicitly shown for clarity. Other diffusion parameter choices 
impact our limits only by up to $\sim$10\%, except for the case of low DM masses, 
for which the effect of solar modulation may be increasingly important 
\cite{DellaTorre:2012zz,Maccione:2012cu}. We reflect this in Fig.~\ref{fig:results} 
by depicting the limits derived in this less certain mass range, where the peak of 
the signal $e^+$ flux (as shown in Fig.~\ref{fig:spectra}) falls below a fiducial value
of 5\,GeV, with dotted rather than solid lines.

For comparison, we have also considered a collection of physical background 
models in which we calculated the expected primary and secondary lepton fluxes 
using GALPROP, and then added the contribution from all galactic pulsars.  While 
this leads to an almost identical description of the background at high energies as 
in the phenomenological model, small differences are manifest at lower energies 
due to solar modulation and a spectral break 
\cite{StrongPapers, Ptuskin:2005ax, Trotta:2010mx} in the CR injection spectrum 
at a few GeV (both neglected in the AMS parameterization). We cross-check our fit
to the AMS positron fraction with lepton measurements by Fermi
\cite{Ackermann:2010ij}. Using these physical background models in our fits, 
instead of the phenomenological AMS parameterization, the limits do not change 
significantly.  The arguably most extreme case would  be the appearance of dips 
in the background due to the superposition of several pulsar contributions, which 
might conspire with a hidden DM signal at almost exactly the same energy. We find 
that in such situations, the real limits on the annihilation rate could be weaker (or 
stronger) by up to roughly a factor of 3 for any individual value of $m_\chi$. See
the Appendix \cite{supp} for more details and further discussion 
of possible systematics that might affect our analysis.

Lastly, we note that the upper limits on $\langle\sigma v\rangle(m_\chi)$
reported in Fig.~\ref{fig:results} can easily be translated into upper limits
on the decay width of a DM particle of mass $2m_\chi$ via $\Gamma \simeq
\langle \sigma v \rangle \rho^\odot_\chi / m_\chi$.  We checked explicitly
that this simple transformation is correct to better than 10\% for the
$L=$4 kpc propagation scenario and $e^+e^-$ and $\mu^+\mu^-$ final states 
over the full considered energy range.

\paragraph*{Conclusions.}

In this Letter, we have considered a possible dark matter contribution to the recent
AMS cosmic ray positron fraction data. The high quality
of this data has allowed us for the first time to successfully perform a
spectral analysis, similar to that used previously in the context of gamma ray 
searches for DM. While we have found no indication of a DM signal, we have
derived upper bounds on annihilation and decay rates into leptonic final
states that improve upon the most stringent current limits by up to two
orders of magnitude.  For light DM in particular, our limits for $e^+e^-$ and 
$\mu^+\mu^-$ final states are significantly below 
the cross section naively predicted for a simple thermal relic. When taken together 
with constraints on DM annihilations to hadronic final states from gamma rays
\cite{Ackermann:2011wa} and antiprotons \cite{Evoli:2011id}, this new information 
significantly limits the range of models which may contain a viable candidate for 
dark matter with $m_\chi\sim\mathcal{O}(10)$\,GeV.

The AMS mission is planned to continue for 20 years. 
Compared to the 18 months of data \cite{Aguilar:2013qda} our analysis is based
on,  we expect to be able to strengthen the presented limits by at least a factor of 
three in the energy range of 6--200\,GeV with the total data set,   and by more in 
the likely case that  systematics and the effective acceptance of the instrument 
improve.

\acknowledgments
This work makes use of SciPy~\cite{SciPy}, Minuit~\cite{James:1975dr} and
Matplotlib~\cite{Hunter:2007}. The research of L.B.~was carried out under
Swedish Research Council (VR) contract no.~621-2009-3915.  T.B.~acknowledges
support from the German Research Foundation (DFG) through the Emmy Noether
grant BR 3954/1-1. I.C., ~C.W.~and D.H.~thank the \emph{Kavli Institute for
Theoretical Physics} in Santa Barbara, California, for their kind hospitality. This 
work has been supported by the US Department of Energy.


\newpage
\section*{\cite{supp} Appendix}

Here, we describe additional tests carried out in order to estimate 
the degree to which our DM limits might vary under alternative assumptions 
pertaining to the astrophysical background and cosmic ray propagation. In 
addition, we quantify the significance of spectral features in the observed 
positron fraction.

\begin{figure*}[!t]
  \includegraphics[width=0.49\linewidth]{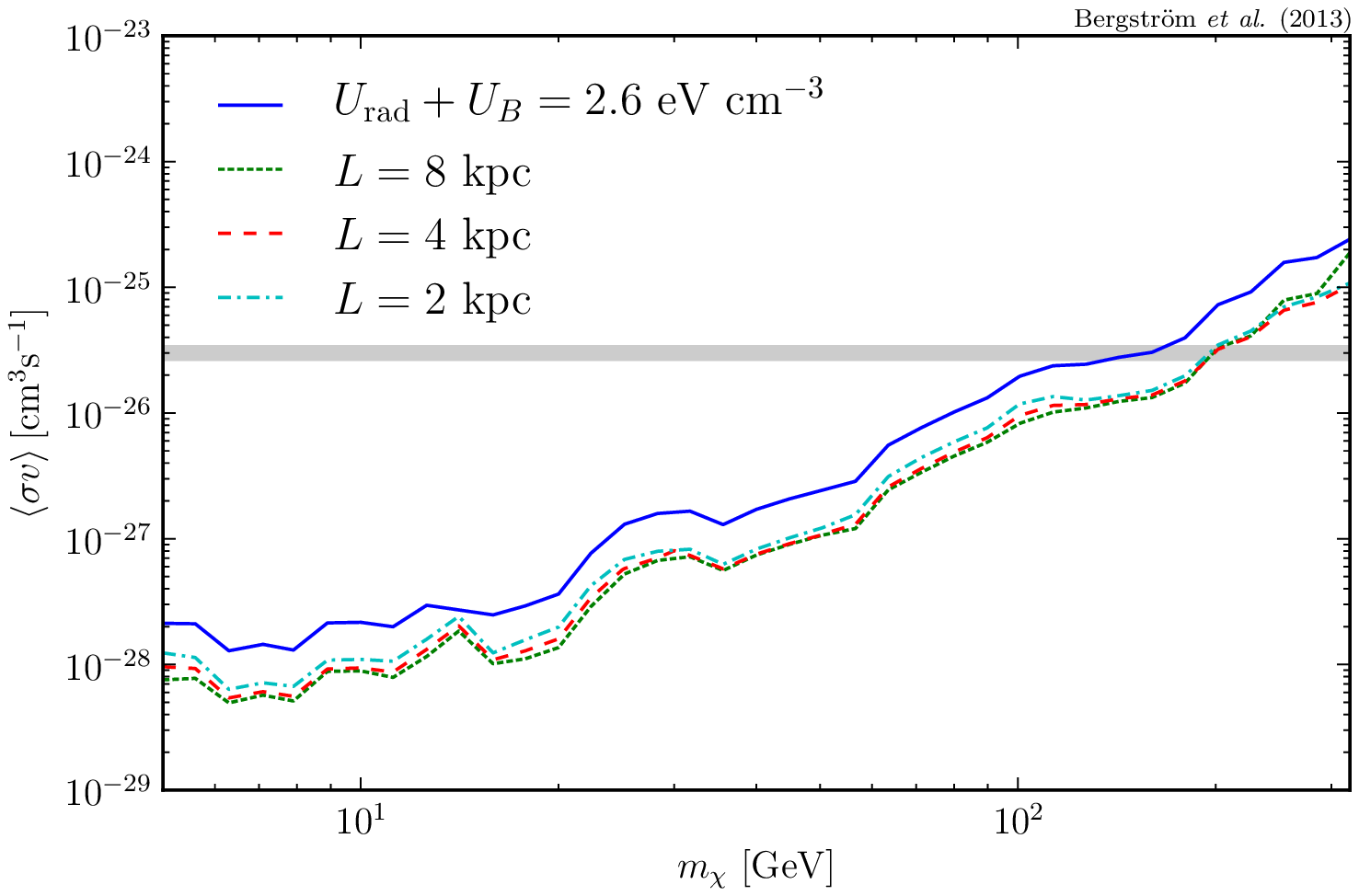}
  \includegraphics[width=0.49\linewidth]{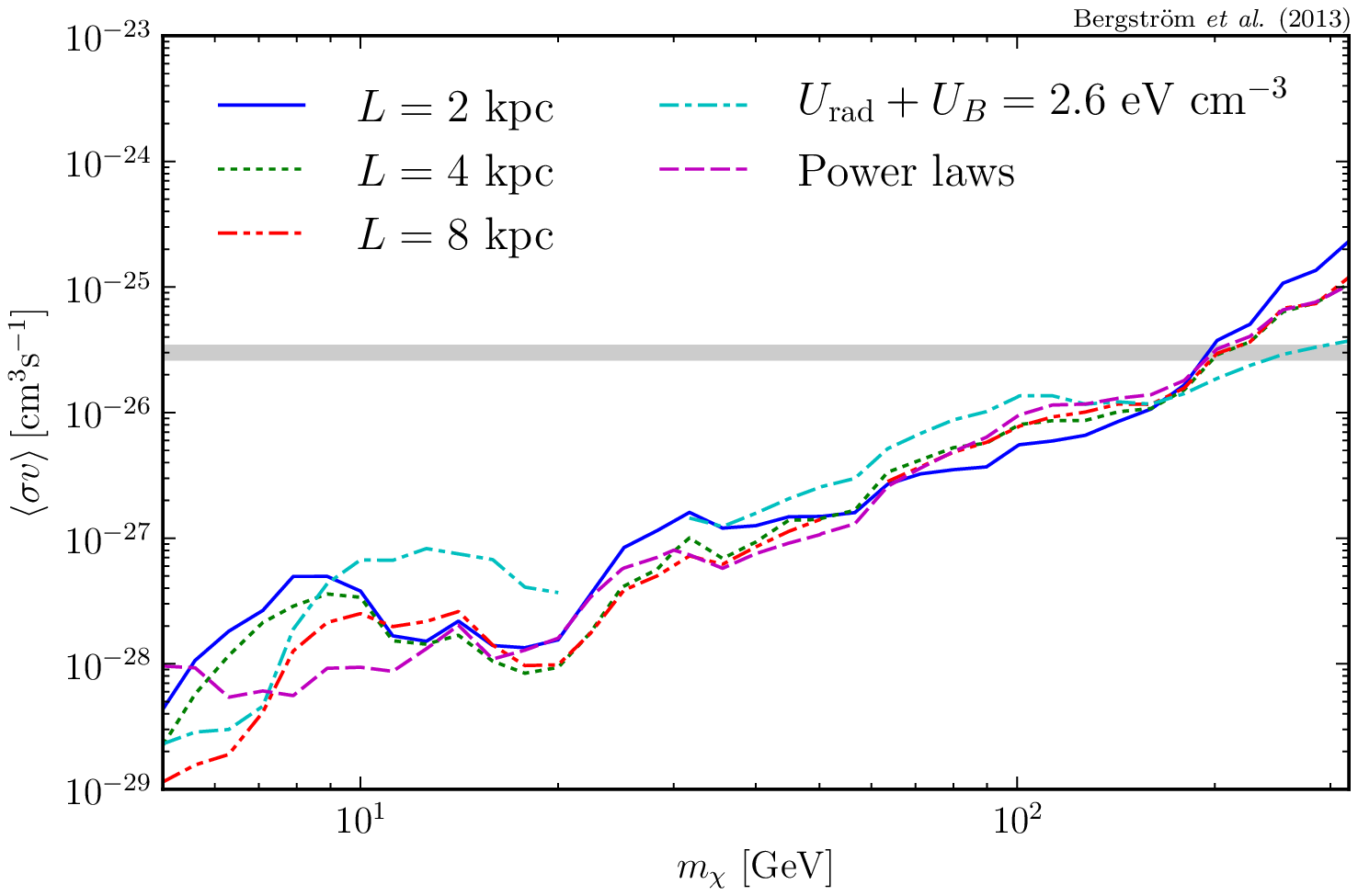}
  \caption{\emph{Left panel:} Limits obtained when different propagation
    models for the DM signal are adopted, using the power-law background 
    model adopted in the main text. \emph{Right panel:} Limits derived using 
    different, physically motivated, background models. In both frames, the 
    results are for the case of DM annihilations to $e^+ e^-$. If not stated
    otherwise, we adopt the benchmark values for $L=4\rm\,kpc$ and the local radiation
    plus magnetic field density $1.7\rm\, eV\,cm^{-3}$.}
  \label{fig:BackPropVariation}
\end{figure*}

In deriving our main results, as shown in Fig.~3, we used the 
phenomenological parameterization of the AMS collaboration 
\cite{Aguilar:2013qda} for the astrophysical contribution to the positron fraction, 
and adopted our reference assumptions of $L=4$ kpc and $U_{rad} + U_{B}
=1.7$ eV cm$^{-3}$.  In Fig.~\ref{fig:BackPropVariation}, for the case of
direct DM annihilation to $e^{+}e^{-}$, we show in the
\textit{left panel} the impact of different propagation parameters when treating 
the astrophysical background in the same way as in Fig.~3. 
Changing the diffusion conditions ($L=2 - 8$ kpc)
in the Galaxy
in that case only affects our limits by $O(10\%)$, while allowing for
higher energy losses ($U_{rad} + U_{B} =2.6$ eV cm$^{-3}$) can alter our
limits by a factor of $\sim$2, with higher losses resulting in weaker limits (see
also Fig.~1).  
In the \textit{right panel}, we repeat this exercise, but replace the AMS background 
parametrization with physically motivated models for the primary $e^{-}$, 
secondary $e^{\pm}$, and pulsar originated $e^{\pm}$ fluxes (see discussion in 
the main text), calculated with the same galactic propagation model as used in 
determining the spectrum of CR leptons from DM. 
In this case, our results can be further altered by a factor of up to $\sim$3. The 
reason for this change is that  our physically motivated models describe the
individual components by power-laws with breaks at a few GeV. These spectral 
features in the background can be the result of different energy loss mechanisms
kicking in,\footnote{
At a few GeV the $e^{\pm}$ energy losses due to
bremsstrahlung emission, dominant at lower energies, equal locally those due to
synchrotron radiation and ICS (dominant at higher energies).  Since the energy
loss rate $dE/dt$ due to bremsstrahlung radiation scales as $E$ while the
$dE/dt$ due to synchrotron and ICS as $E^2$ (at the Thompson cross-section
regime), a spectral change in the propagated $e^{\pm}$ around that energy is
expected (see, e.g., Ref…~\cite{Bringmann:2011py}).} 
or from individual local and recent supernovae affecting the high energy
$e^{-}$ spectrum. Also, observations at microwave and radio frequencies 
suggest a different spectral power-law for the CR $e^{\pm}$ at $\sim$1\,GeV  
\cite{StrongPapers,Bringmann:2011py,Jaffe:2011qw} 
compared to CR $e^{\pm}$ flux measurements at higher energies   
\cite{FermiLAT:2011ab, Adriani:2011xv}.
While changes in the spectral power-law describing these
components are motivated by the reasons just described, sharp breaks used to
implement them are theoretically less accurate and fit slightly worse the AMS
positron fraction spectrum.

In addition, our physically motivated models include the impact of solar
modulation by using the force field approximation. Solar modulation modifies the
position and normalization of the dark matter signal flux, but is negligible at 
energies $>$5\,GeV.\footnote{
In certain models, solar modulation can also
affect the observed height of the peak in the positron fraction by changing the 
ratio of electrons-to-positrons of same energy before entering the Heliosphere 
\cite{Strauss2012, Strauss2012B}.} 
We do not expect solar modulation to 
significantly smoothen a sharp spectral peak at higher energies.  \medskip

Given that we consider a population of pulsars as one possible source 
of the rising positron fraction above 10\,GeV (with TeV-scale DM or a single 
dominant  pulsar being alternative possibilities), we will briefly discuss the 
impact of their modeling on our limits. For pulsars that eventually inject equal 
amounts of $e^{\pm}$ 
into the ISM, their injection spectra can be estimated from gamma-ray and 
synchrotron observations towards known pulsars, such as the Crab.\footnote{
Yet the
uncertainties are still large due to a lack of exact understanding of local
environments or the type of relevant supernova remnants (within which the pulsars 
exist); typically, the $e^\pm$ also get further accelerated at the termination shock 
between the magnetosphere and the pulsar wind nebulae (PWN), and the $e^{\pm}$ 
injected to the ISM are dominantly coming from middle 
aged pulsars after their respective PWN have been disrupted 
(see Refs.~\cite{Arons, Green:2009qf, Malyshev:2009tw}) } 
Typical injection power-law values for the differential spectrum  are expected 
to be in the range of 1-2  leading to propagated spectra with power-laws in the 
range of 2.0 $\pm$ 0.5. Our fits for the averaged pulsar contribution agree
with these expectations.

\begin{figure}[t]
  \begin{center}
  \includegraphics[width=\linewidth]{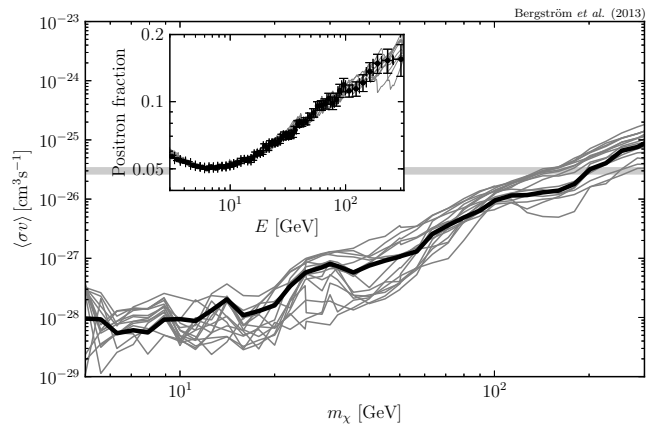}
  \end{center}
  \caption{ The \emph{black line} shows our nominal limit on $e^+e^-$ final
    states, obtained by adopting the power-law background model. The 
    \emph{gray lines}, in contrast, show limits obtained when the contribution 
    from many pulsars is taken into account (for 15 different realizations).}
  \label{fig:LimitsRand}
\end{figure}

In addition, as suggested by Refs.~\cite{Kobayashi:2003kp, Malyshev:2009tw,
Grasso:2009ma}, the total contribution from many pulsars -- each with a 
different age, distance, initial rotational energy, injected energy into $e^{\pm}$, 
and unique environmental surroundings affecting energy losses and diffusion -- 
is  expected to give a spectrum with many peaks and dips, especially at higher
energies where fewer pulsars significantly contribute. With fine enough energy 
resolution and high statistics, one should be able to observe such spectral 
features. By using the data from the ATNF pulsar catalogue \cite{ATNF} and 
implementing the parametrization of Ref.~\cite{Malyshev:2009tw}, we ran 
multiple realizations of such combined spectra to study the impact of possible 
dips and peaks in the background spectrum on the derived DM limits.  
In particular, we include in these realizations all pulsars within 4\,kpc from us, 
except for millisecond pulsars and pulsars in binary systems. While 
we keep their individual locations and ages in all realizations fixed, we vary 
i) the local CR diffusion properties and energy-losses, ii) the cuts on the current 
spin-down power of pulsars as recorded in the ATNF catalogue 
and, most importantly, iii) the fraction $\eta$ of initial rotational energy of the 
individual neutron stars that is injected into the ISM in the form of $e^{\pm}$. 

We then fit to the AMS data the injection spectral properties
(taken to be the same for all pulsars), the \textit{averaged} value of
$\eta$, the primary SNe $e^{-}$, secondary $e^{\pm}$ CR flux 
normalizations and the solar modulation potential. 
Even though the injection $e^{\pm}$ spectrum is taken to be the 
same for simplicity, however, the propagated pulsar spectra differ because of
 the different ages, distances and energy outputs; this is clearly seen in the
 inset of Fig.~\ref{fig:LimitsRand} where we plot the resulting positron fraction.
For DM 
channels that give broad continuous spectra, such as muons and taus (see 
Fig.~1), the presence of multiple peaks and dips is unimportant. For 
hard spectra such as from monochromatic $e^{\pm}$, however, our limits can 
be modified by a factor of up to $\sim$3, as also shown in Fig.~\ref{fig:LimitsRand}.

\begin{figure}[t]
  \includegraphics[width=\linewidth]{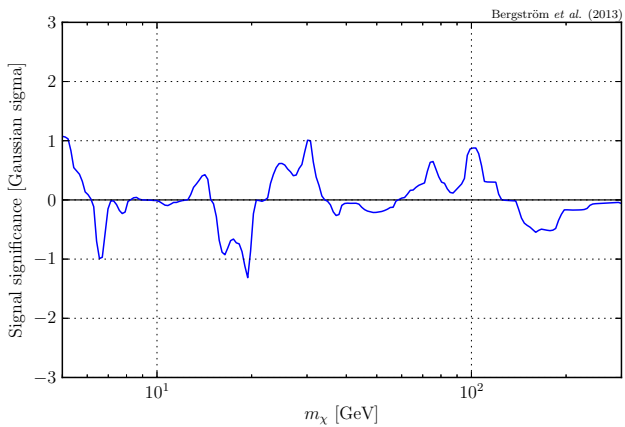}
  \caption{Significance for a contribution from a $e^+e^-$ DM signal to the
  AMS-02 positron fraction, for different DM energies, in units of Gaussian
  sigma. Negative values correspond to negative (but unphysical) signal
  normalizations.}
  \label{fig:TS}
\end{figure}

In Fig.~\ref{fig:TS} we show, for the case of $e^+e^-$ final states, the local
significance for a DM signal as function of the DM mass. The significance is
plotted in units of Gaussian sigma, and given by the square-root $\sqrt{TS}$
of the Test Statistics $TS=-2\log \mathcal{L}_{\rm null}/\mathcal{L}_{\rm
alt}$. Here, $\mathcal{L}_{\rm alt/null}$ denote respectively the likelihood
of the alternative (DM signal) and null (no DM signal) hypothesis. For
illustration, we also allow negative (obviously unphysical) signal
normalizations in the fit, which are mapped onto the negative y-axis. As
background model in the fit we use the reference power-law model
from~Ref.~\cite{Aguilar:2013qda}. We do not find any indications for local, 
edge-like, features in the AMS data. 

Lastly, as a simple cross-check, we have also run DarkSUSY
\cite{Gondolo:2004sc} with standard parameters for propagation  (based on
the prescription given in Ref.~\cite{Baltz:1998xv}) and an NFW 
profile normalized to 0.4 GeV cm$^{-3}$. For the electron spectrum, we 
used a simple broken power law
which agrees with the PAMELA electron data \cite{Adriani:2011xv} for $E>5$\,GeV.
Knowing that the AMS positron fraction measurement is well described by a simple 
background model,  we then just demand that the DM signal does not exceed
the reported $2\sigma$ error bars at  the energy 
of the feature. The resulting limit curve agrees well with the more sophisticated 
treatment described in the main text.

\end{document}